\documentclass[12pt]{article}
\usepackage{amssym}
\usepackage{graphics}

\newcommand{\htt}{\hat{T}}
\newcommand{\hp}{\hat{p}}
\def\R{\mbox{$\Bbb R$}}

\def\sech{\mathop{\rm sech}\nolimits}
\def\case#1#2{{\textstyle{#1\over #2}}}

\title{
A GENERAL SCHEME FOR THE EFFECTIVE-MASS SCHR\"ODINGER EQUATION AND THE
GENERATION OF THE ASSOCIATED POTENTIALS}
\author{B. BAGCHI$^*$ and P. GORAIN\\
{\small \sl Department of Applied Mathematics, University of Calcutta,} \\
{\small \sl 92 Acharya Prafulla Chandra Road, Kolkata 700 009, India}\\
{\small \sl $^*$bbagchi123@rediffmail.com}\\ [7pt] 
C. QUESNE\\
{\small \sl Physique Nucl\'eaire Th\'eorique et Physique Math\'ematique,}
\\ {\small \sl Universit\'e Libre de Bruxelles, Campus de la Plaine CP229,} \\ 
{\small \sl Boulevard~du Triomphe, B-1050 Brussels, Belgium} \\
{\small \sl cquesne@ulb.ac.be}\\[7pt]
R. ROYCHOUDHURY\\
{\small \sl Physics and Applied Mathematics Unit, Indian Statistical Institute,}\\
{\small \sl Kolkata 700 035, India}\\
{\small \sl raj@isical.ac.in}}
\date{ }
\begin{document}
\baselineskip=22pt plus 1pt minus 1pt
\maketitle

\begin{abstract} 
A systematic procedure to study one-dimensional Schr\"odinger equation with a
position-dependent effective mass (PDEM) in the kinetic energy operator is explored. The
conventional free-particle problem reveals a new and interesting situation in that, in the
presence of a mass background, formation of bound states is signalled. We also discuss 
coordinate-transformed, constant-mass Schr\"odinger equation, its matching with the
PDEM form and the consequent decoupling of the ambiguity parameters. This provides
a unified approach to many exact results known in the literature, as well as to a lot of new
ones.
\end{abstract}

\noindent
Running head: Effective-Mass Schr\"odinger Equation


%
%
\newpage
\section{Introduction}

In recent times, the concept of a position-dependent-effective-mass (PDEM)
quantum Hamiltonian is rapidly gaining acceptance because of its increasing relevance in
describing the motion of electrons in problems of compositionally graded
crystals~\cite{geller} (following our ability to fabricate semiconductor nanostructures),
quantum dots~\cite{serra}, liquid crystals~\cite{barranco}, etc. The appearance
of PDEM is also well known in the energy density functional approach to the nuclear
many-body problem~\cite{ring} and its applications~\cite{arias, puente} in the context of
nonlocal terms of the accompanying potential. Other theoretical considerations where
PDEM has been exploited include the derivation~\cite{levy} of the underlying electron
Hamiltonian from instantaneous Galilean invariance and implementation of the path integral
techniques~\cite{yung} to calculate the Green's function~\cite{chetouani} for step and
rectangular-barrier potentials and masses. Further, PDEM has proved to be appealing in
the construction of acceptable quantum mechanical systems by seeking exact solutions of
the Schr\"odinger equation~\cite{dekar98, dekar99, plastino00, koc02, gonul02a,
alhaidari, dutra00, dutra03,yu} by extending the already existing methods of spectrum
generating or potential algebras~\cite{roy} and those of supersymmetric quantum
mechanics~\cite{milanovic, plastino98, gonul02b, cq}.\par
%
%
In this Letter we consider the PDEM formalism to address the issue of the generation
of the corresponding potentials. We use the von Roos effective-mass kinetic energy
operator~\cite{vonroos}, which has the advantage of an inbuilt Hermiticity, and we
develop a PDEM scheme in very general terms that not only encompasses some of the
previous ones but also brings out additional features not reported before.\par
%
%
{}First, we consider the problem of a free particle in the framework of a first-order
intertwining relationship and show that in a suitable mass background it acts as a bound
system.\par
%
%
Second, we observe that a Schr\"odinger equation with constant mass and having for its
potential $U$ can always be coordinate transformed in terms of some given mass function
such that the resulting form matches with the one implied by PDEM. In this way the
starting potential $V$ of the PDEM scheme becomes expressible in terms of the
coordinate-transformed $U$ and the one induced by the effective kinetic energy operator
$\htt$. The latter is found to depend on two types of parameters: those entering the
definition of the mass function and those coming from the ordering ambiguity of the
momentum and mass operators.\par
%
%
Third, although it is clear that $\htt$ provides a nontrivial contribution to the effective
potential which a particle with a position-dependent mass experiences, it is not at all
apparent how $V$ stands out against the interplay of these ambiguity parameters
vis-\`a-vis the ones of the coordinate-transformed $U$. In fact, as we shall see, in a
PDEM problem the ambiguity parameters get decoupled in a natural way. This observation
is new and of significance in that it allows $V$ to become entirely identifiable with the
coordinate-transformed $U$. The latter in the constant-mass limiting case goes over to a
known form in the sense that it is either exactly solvable (ES) or quasi-exactly solvable
(QES) or else conditionally exactly solvable (CES). We also discuss some
consequences of our results.\par
%
%
\section{General Strategy and the Free-particle Problem}

The most commonly used effective-mass kinetic energy operator is the
two-parameter form given by von Roos~\cite{vonroos}, which in one dimension reads
\begin{equation}
  \htt = \frac{1}{4}\left[m^{\alpha}(x)\, \hp\, m^{\beta}(x)\, \hp\, m^{\gamma}(x) +
  m^{\gamma}(x)\, \hp\, m^{\beta}(x)\, \hp\, m^{\alpha}(x)\right],  \label{eq:T}
\end{equation}
where $\hp$ represents the momentum operator, $m(x)$ is the position-dependent
mass and the parameters $\alpha$, $\beta$, $\gamma$ are subject to the condition
$\alpha + \beta + \gamma = -1$. Because of the noncommutativity of the momentum
and (position-dependent) mass operators, maintaining Hermiticity of the kinetic energy
operator is not trivial. In (\ref{eq:T}), $\htt$ has been specifically designed to be
Hermitian but is by no means unique. However, it turns out that other plausible forms of
$\htt$ invariably reduce to one of the special cases of
(\ref{eq:T}), apart from the fact that most of them are equivalent~\cite{geller} up to a
given level of accuracy.\par
%
%
On the other hand, analyses have been performed comparing experimental data with
theoretical insights provided by various considerations such as, for example, the study of
the envelope function for electrons in uniform or slowly graded crystals~\cite{geller}.
These, however, do not always reveal~\cite{morrow} precise and unambiguous estimates
to the question of uniqueness of the parameters $\alpha$, $\beta$, $\gamma$. Some of
the appropriate single-band PDEM Hamiltonians are the ones of BenDaniel and
Duke (BDD)~\cite{bendaniel} ($\alpha = 0$, $\beta = -1$), Bastard~\cite{bastard}
($\alpha = -1$, $\beta = 0$), Zhu and Kroemer (ZK)~\cite{zhu} ($\alpha = -
\frac{1}{2}$, $\beta = 0$) and the redistributed model~\cite{li} ($\alpha =0$, $\beta = -
\frac{1}{2}$).\par
%
%
With the correspondence $\hp \to - {\rm i}\hbar \frac{d}{dx}$ in (\ref{eq:T}), the
time-independent Schr\"odinger equation reads
\begin{eqnarray}
  &&\left\{- \frac{\hbar^2}{4} \left[m^{\alpha}(x) \frac{d}{dx} m^{\beta}(x)
        \frac{d}{dx} m^{\gamma}(x) + m^{\gamma}(x) \frac{d}{dx} m^{\beta}(x)
        \frac{d}{dx} m^{\alpha}(x)\right] + V(x)\right\} \psi(x) \nonumber \\
&& \qquad =E \psi(x)  \label{eq:SE-PDEM}
\end{eqnarray}
for some given potential $V(x)$. On setting
\begin{equation}
  m(x) = m_0 M(x),
\end{equation}
where $M(x)$ is the dimensionless form of the mass function, along with $\hbar = 2m_0
= 1$, we can get rid of the ambiguity parameters by transferring them to the effective
potential energy of the variable-mass system. Thus using the result
\begin{eqnarray}
  \lefteqn{M^{\alpha} \frac{d}{dx} M^{\beta} \frac{d}{dx} M^{\gamma} +
        M^{\gamma} \frac{d}{dx} M^{\beta} \frac{d}{dx} M^{\alpha}} \nonumber \\
  & = & 2 \frac{d}{dx} \frac{1}{M} \frac{d}{dx} - (\beta+1) \frac{M''}{M^2} + 2 [\alpha
        (\alpha+\beta+1) + \beta + 1] \frac{M^{\prime2}}{M^3}, 
\end{eqnarray}
where a prime denotes derivative with respect to $x$, equation~(\ref{eq:SE-PDEM})
acquires the form
\begin{equation}
  H \psi(x) \equiv \left[- \frac{d}{dx} \frac{1}{M(x)} \frac{d}{dx} + V_{\rm eff}(x)\right]
  \psi(x) = E \psi(x),  \label{eq:SE-PDEM-ter}
\end{equation}
in which the effective potential $V_{\rm eff}(x)$ is seen to depend on some mass terms:
\begin{equation}
  V_{\rm eff}(x) = V(x) + \frac{1}{2} (\beta+1) \frac{M''}{M^2} - [\alpha
        (\alpha+\beta+1) + \beta + 1] \frac{M^{\prime2}}{M^3}.  \label{eq:Veff} 
\end{equation}
In the following, we shall be interested in bound-state eigenvalues $E_n$, $n=0$, 1,
2,~\ldots, and corresponding wavefunctions $\psi_n(x)$, $n=0$, 1, 2,~\ldots.\par
%
%
Let us consider the intertwining relationship
\begin{equation}
  \eta H = H_1 \eta,  \label{eq:intertwine}
\end{equation}
where $H_1$ has the same kinetic energy term as $H$ and an associated potential
$V_{1,{\rm eff}}(x)$. If the ground-state wavefunction $\psi_0$ of $H$ is annihilated by
the operator
$\eta$, i.e.,
$\eta \psi_0 = 0$, the eigenvalues of $H_1$ are $E_{1,n} = E_{n+1}$, $n=0$, 1,
2,~\ldots, with corresponding wavefunctions $\eta \psi_{n+1}$, since from
(\ref{eq:intertwine}) it follows that  $H_1 (\eta \psi_{n+1}) = \eta H \psi_{n+1} =
E_{n+1} (\eta \psi_{n+1})$ for $n=0$, 1, 2,~\ldots.\par
%
%
Choosing a first-derivative intertwining operator $\eta = A(x) \frac{d}{dx} + B(x)$ in
(\ref{eq:intertwine}), we are led to the restrictions
\begin{eqnarray}
  A(x) & = & M^{-1/2},  \label{eq:A} \\
  V_{\rm eff}(x) & = & \lambda + B^2 - (AB)', \label{eq:Veff-int}\\
  V_{1, {\rm eff}}(x) & = & V_{\rm eff} + 2AB' - AA'', 
\end{eqnarray}
where $\lambda$ denotes some integration constant.\par
%
%
It is instructive to consider the free-particle case $V(x) = V_0$ of (\ref{eq:Veff}). On
comparing with (\ref{eq:Veff-int}), where we choose $\lambda = V_0$ and
\begin{equation}
  B(x) = - \frac{1}{2} (\beta+1) \frac{M'}{M^{3/2}},  \label{eq:B}
\end{equation}
we get a constraint relation $\beta = - 2\alpha - 1$ and equation~(\ref{eq:Veff}) is found
to be consistent with the PDEM Hamiltonians of BDD ($\alpha=0$, $\beta=-1$) and ZK
($\alpha = - \frac{1}{2}$, $\beta=0$). Specifically, $V_{\rm eff}$ and $V_{1,{\rm
eff}}$ read
\begin{eqnarray}
  V_{\rm eff}(x) & = & V_0 - \alpha \frac{M''}{M^2} + \alpha(\alpha+2)
        \frac{M^{\prime2}}{M^3}, \\
  V_{1,{\rm eff}}(x) & = & V_0 + \left(\alpha + \frac{1}{2}\right) \frac{M''}{M^2} +
        \left(\alpha + \frac{1}{2}\right) \left(\alpha - \frac{3}{2}\right)
        \frac{M^{\prime2}}{M^3}, 
\end{eqnarray}
which are interchanged under the transformation $\alpha \to - (\alpha +
\frac{1}{2})$. In particular, for $\alpha=0$,
\begin{equation}
  V_{\rm eff}(x) = V_0, \qquad V_{1,{\rm eff}}(x) = V_0 + \frac{1}{2}
        \frac{M''}{M^2} - \frac{3}{4} \frac{M^{\prime2}}{M^3}, 
\end{equation}
while for $\alpha = - \frac{1}{2}$,
\begin{equation}
  V_{\rm eff}(x) = V_0 + \frac{1}{2} \frac{M''}{M^2} - \frac{3}{4}
        \frac{M^{\prime2}}{M^3}, \qquad V_{1,\rm eff}(x) = V_0,  \label{eq:ZK}
\end{equation}
suggesting a duality between the BDD and ZK schemes. It is to be stressed that the
results (\ref{eq:A}) -- (\ref{eq:ZK}) are independent of any choice of $M(x)$.\par
%
%
To proceed further with the effective-mass Schr\"odinger equation
(\ref{eq:SE-PDEM-ter}) we need to have a precise example for $M(x)$. We can
take for instance a deformed hyperbolic function
\begin{equation}
  M(x) = \sech^2 qx, \qquad q>0,  \label{eq:mass1}
\end{equation}
which depicts a solitonic profile. Another acceptable form for $M(x)$ is
\begin{equation}
  M(x) = \left(1 + \frac{q}{1+x^2}\right)^2, \qquad q>0,  \label{eq:mass2}
\end{equation}
which has yielded interesting connections~\cite{roy} with the su(1,1) algebra. In
(\ref{eq:mass1}) or (\ref{eq:mass2}), $q$ may be treated as a deformation parameter
so that when $q \to 0$, $M(x) \to 1$ in both the cases.\par
%
%
Let us take for concreteness the result (\ref{eq:ZK}), which is in conformity with the ZK
scheme. We can write for the counterpart of (\ref{eq:SE-PDEM-ter}) for $H_1$
\begin{equation}
  - \frac{d}{dx} \left(\frac{1}{M} \frac{d\varphi_n}{dx}\right) = E'_{1,n} \varphi_n,
  \label{eq:H_1}
\end{equation}
where $E'_{1,n} = E_{1,n} - V_0 = E_{n+1} - V_0 = E'_{n+1}$ and $\varphi_n(x) = \eta
\psi_{n+1}(x)$ are the eigenfunctions of $H_1$. Using the specific example
(\ref{eq:mass1}) for $M(x)$ and taking (\ref{eq:A}) and (\ref{eq:B}) into account, we
can write these eigenfunctions as $\varphi_n(x) = \frac{d}{dx}(\cosh qx\, \psi_{n+1})$.
Inserting this expression in (\ref{eq:H_1}), integrating once the resulting equation,
changing $n+1$ into $n$ and rewriting the result in terms of $\chi_n(t) = \cosh qx\,
\psi_n(x)$, where
$t =
\tanh qx$, we find
\begin{equation}
  (1 - t^2) \frac{d^2 \chi_n}{dt^2} - 2t \frac{d\chi_n}{dt} + \frac{E'_n}{q^2} \chi_n
  = 0.  \label{eq:Legendre}
\end{equation}
Such an equation coincides with the equation for Legendre polynomials $\chi_n(t) =
P_n(t)$, $n=0$, 1, 2,~\ldots, provided $E'_n = q^2 n(n+1)$.\footnote{Actually the
general solutions ${}_2F_1(-\nu, \nu+1; 1; (1 - \tanh qx)/2)$ of Legendre's differential
equation (\ref{eq:Legendre}), corresponding to $E'_{\nu} = q^2 \nu(\nu+1)$ ($\nu
\ne n$), being not convergent for $x \to - \infty$, the polynomial solutions are the only
acceptable ones.} From this, it follows that apart from some normalization factors,
$\psi_n(x) \propto \sech qx P_n(\tanh qx)$ and $\varphi_n(x) \propto \frac{d}{dx}
P_{n+1}(\tanh qx)$. We therefore get for the first few normalizable solutions of $H$
and $H_1$,
\begin{eqnarray}
  \psi_0(x) & \propto & \sech qx, \qquad E'_0 = 0, \nonumber \\
  \psi_1(x) & \propto & \sech qx \tanh qx, \qquad \varphi_0(x) \propto \sech^2 qx,
        \qquad E'_1 = E'_{1,0} = 2q^2, \nonumber \\
  \psi_2(x) & \propto & \sech qx (1 - \case{3}{2}\sech^2 qx), \qquad \varphi_1(x)
        \propto \sech^2 qx \tanh qx, \nonumber \\
  && E'_2 = E'_{1,1} = 6q^2, \nonumber \\
  \psi_3(x) & \propto & \sech qx \tanh qx (1 - \case{5}{2}\sech^2 qx), \qquad
        \varphi_2(x) \propto \sech^2 qx (1 - \case{5}{4}\sech^2 qx), \nonumber \\
  && E'_3 = E'_{1,2} = 12q^2, \nonumber \\
  \psi_4(x) & \propto & \sech qx (1 - 5 \sech^2 qx + \case{35}{8} \sech^4 qx),
        \qquad \varphi_3(x) \propto \sech^2 qx \tanh qx \nonumber \\
  && \mbox{} \times (1 - \case{7}{4}\sech^2 qx), \qquad E'_4 = E'_{1,3} = 20q^2. 
\end{eqnarray}  
We conclude that a free particle placed in an appropriate mass background generates
bound states in a manner as given above.\par
%
%
\section{Coordinate Transformation}

Equation (\ref{eq:SE-PDEM-ter}) can be interpreted as a coordinate-transformed
Schr\"odinger equation with the coordinate dependence arising from a suitable choice of
mass function $M(x)$. To this end, let us consider a time-independent Schr\"odinger
equation with constant mass under the action of a potential $U$,
\begin{equation}
  \left[- \frac{d^2}{dy^2} + U(y;a)\right] \phi(y) = \epsilon \phi(y), 
  \label{eq:SE-constant}
\end{equation}
where $a$ denotes collectively a set of coupling parameters, which may be present in
$U$, and $\epsilon$ is the energy.\par
%
%
The change of variable
\begin{eqnarray}
  y & = & \lambda z(x) + \nu, \qquad \lambda, \nu \in \R,  \label{eq:y} \\
  z(x) & = & \int^x \sqrt{M(x')}\, dx',  \label{eq:mu}
\end{eqnarray}
transforms equation~(\ref{eq:SE-constant}) into
\begin{equation}
  \left[- \frac{1}{\sqrt{M}} \frac{d}{dx} \frac{1}{\sqrt{M}} \frac{d}{dx} + \lambda^2
  U\Bigl(\lambda z(x) + \nu; a\Bigr)\right] \chi(x) = \lambda^2 \epsilon \chi(x),
  \label{eq:SE-transf}
\end{equation}
where $\chi(x) \equiv \phi(y(x))$.\par
%
%
To compare (\ref{eq:SE-transf}) with (\ref{eq:SE-PDEM-ter}), one needs to write the
term containing the second-order derivative in the former to transform in the same way
as in the latter. For such a purpose, we exploit the property $\frac{1}{\sqrt{M}}
\frac{d}{dx} = \frac{d}{dx} \frac{1}{\sqrt{M}}+ \frac{1}{2} \frac{M'}{M^{3/2}}$ and
substitute $\chi(x) = M^{-1/4}(x) \psi(x)$  to reset (\ref{eq:SE-transf}) as
\begin{equation}
  \left[- \frac{d}{dx} \frac{1}{M} \frac{d}{dx} + \frac{M''}{4M^2} - \frac{7M^{\prime2}}
  {16M^3} + \lambda^2 U\Bigl(\lambda z(x) + \nu; a\Bigr)\right] \psi(x) = \lambda^2
  \epsilon \psi(x).  \label{eq:SE-transf-ter}
\end{equation}
\par
%
%
Equations (\ref{eq:SE-PDEM-ter}) and (\ref{eq:SE-transf-ter}) are seen to coincide (with
$E = \lambda^2 \epsilon$) provided $V_{\rm eff}$ is identified as
\begin{equation}
  V_{\rm eff}(x) = \lambda^2 U\Bigl(\lambda z(x) + \nu; a\Bigr) + \frac{M''}{4M^2} 
  - \frac{7M^{\prime2}}{16M^3}.  \label{eq:Veff-bis}
\end{equation}
From (\ref{eq:Veff}) and (\ref{eq:Veff-bis}), we thus arrive at an important result
\begin{equation}
  V(x) = V_1(x; \alpha, \beta) + V_2(x; a),  \label{eq:V}
\end{equation}
where $V_1(x; \alpha, \beta)$ and $V_2(x; a)$ are
\begin{eqnarray}
  V_1(x; \alpha, \beta) & = & \left[\alpha(\alpha+\beta+1) + \beta + \frac{9}{16}\right]
        \frac{M^{\prime2}}{M^3} - \frac{1}{4} (2\beta+1) \frac{M''}{M^2}, \\
  V_2(x; a) & = & \lambda^2 U\Bigl(\lambda z(x) + \nu; a\Bigr).  \label{eq:V2}
\end{eqnarray} 
\par
%
%
We see at once that for any solution $(\epsilon, \phi(y))$ of the constant-mass
Schr\"odinger equation, we can derive a corresponding one $(E, \psi(x))$ of the
effective-mass Schr\"odinger equation with $E = \lambda^2 \epsilon$ and the potential
$V(x)$ given by the set (\ref{eq:V})~--~(\ref{eq:V2}).\par
%
%
\section{\boldmath Analysis of $V(x)$}

Of the two parts of $V(x)$, one involves the ambiguity parameters of the $\htt$
operator, which we call $V_1(x; \alpha, \beta)$, while the other (i.e.\ $V_2(x; a)$) is a
coordinate-transformed piece obtained as a result of coordinate transforming $U$
according to (\ref{eq:y}) and (\ref{eq:mu}).\par
%
%
\subsection{\boldmath The potential $V_1(x; \alpha, \beta)$}

Unrestricted $\alpha$, $\beta$ parameters are often of much use to study the general
class of solutions for the PDEM Schr\"odinger equation and to understand their role when
the system undergoes a transition from a smooth potential and mass step to an abrupt
situation~\cite{dekar99}. In spite of this, an interesting question is whether the presence
of $V_1$ can be eliminated from $V$ for special choices of $\alpha$, $\beta$ and/or the
mass function $M(x)$. The answer turns out to be in the affirmative as our analysis below
will presently confirm.\par
%
%
Writing $V_1(x; \alpha, \beta)$ as 
\begin{equation}
  V_1(x; \alpha, \beta) = f(\alpha, \beta) \frac{M^{\prime2}}{M^3} - g(\beta) 
  \frac{M''}{M^2},
\end{equation}
where $f(\alpha, \beta) = \left(\alpha + \frac{1}{4}\right)^2 + \frac{1}{2} (\alpha+1)
(2\beta+1)$, $g(\beta) = \frac{1}{4} (2\beta+1)$, we notice that $V_1$ vanishes for $
\beta = - \frac{1}{2}$ and $\alpha = - \frac{1}{4}$, corresponding to any smooth mass
function $M(x)$. The value $\beta = - \frac{1}{2}$ is consistent with the
redistributed model~\cite{li}, which is a special case of (\ref{eq:T}), as noted earlier.
Although the accompanying zero value of $\alpha$ is different from $- \frac{1}{4}$, the
PDEM kinetic energy operators are known~\cite{ribeiro, cavalcante} to be
equivalent over a wide range of $\alpha$ values for suitable width of the transition
region.\par
%
%
{}For $\beta \ne - \frac{1}{2}$, the vanishing of $V_1(x; \alpha, \beta)$ also takes
place for the following mass dependence
\begin{eqnarray}
  && f \ne g: \qquad M \sim x^{\xi}, \qquad \xi = \left(1 - \frac{f}{g}\right)^{-1}, \\
  && f = g: \qquad M \sim e^{\pm kx}, \qquad k \in \R^+,
\end{eqnarray}
i.e., either for power and inverse power laws or for exponentially rising and falling masses.
For the kinetic energy operators of BDD~\cite{bendaniel}, Bastard~\cite{bastard}, and
ZK~\cite{zhu}, $\xi$ turns out to be $- \frac{4}{3}$, $- \frac{4}{5}$, $-4$, respectively.
In such cases, $M(x)$ is singular for $x \to 0$. In general (i.e., notwithstanding
the vanishing of $V_1$), the case $f = g$ is disfavoured by the above schemes because
the underlying constraint equation $\left(\alpha + \frac{1}{4}\right)^2 + \frac{1}{2}
\left(\alpha + \frac{1}{2}\right) (2\beta+1) = 0$ is not satisfied by the corresponding
parameter values. However, exponential dependence on the position coordinate of the
mass has been considered~\cite{gonul02b} viable in semiconductor quantum well
structures and also in problems of exact solvability in the supersymmetric context.\par
%
%
\subsection{\boldmath The potential $V_2(x; a)$}

The comprehensive nature of our scheme enables us to obtain rather straightforwardly
the relevant forms of $V_2$ should $U$ for the constant-mass Schr\"odinger equation
be available. To show how our scheme works in practice, we consider the following
illustration of the Scarf I potential, which is well known to be ES on the interval
$\left(- \frac{\pi}{2}, \frac{\pi}{2}\right)$:
\begin{eqnarray}
  U(x; A, B) & = & [A(A-1) + B^2] \sec^2x - B (2A-1) \sec x \tan x - A^2, 
        \nonumber  \label{eq:Scarf} \\
  && 0 < B < A-1, \\
  \epsilon_n & = & (A+n)^2 - A^2, \qquad n = 0, 1, 2, \ldots.  \label{eq:Scarf-E}
\end{eqnarray}
\par
%
%
Using (\ref{eq:mu}), (\ref{eq:V2}) and say, for instance, (\ref{eq:mass2}), we easily find
that in the presence of mass deformation the argument $x$ in (\ref{eq:Scarf}) is modified
according to $x \to \lambda z(x) + \nu$, $z(x) \equiv x + q \tan^{-1}x$. As a result,
$V_2(x; A, B)$ can be written as
\begin{equation}
  V_2(x; A, B) = [A(A-1) + B^2] \sec^2[z(x)] - B (2A-1) \sec[z(x)] \tan[z(x)] - A^2, 
  \label{eq:Scarf-q}
\end{equation}
where along with $\nu=0$, we have set $\lambda=1$ so that the energy levels are the
same as given by (\ref{eq:Scarf-E}), i.e., $E_n = \epsilon_n$.\par
%
%
A graphical representation of (\ref{eq:Scarf-q}) is shown in figure~1, where we have
considered the effects of $q$-deformed mass function (\ref{eq:mass2}) for various $q$
values and compared them with the undeformed (i.e. $q=0$) case (\ref{eq:Scarf}). It is
to be noticed that $V_2(x; A, B)$ is not singular provided
\begin{equation}
  - \frac{1}{q} \left(\frac{\pi}{2} + x\right) < \tan^{-1}x < \frac{1}{q}
  \left(\frac{\pi}{2} - x\right). 
\end{equation}
Hence it is defined on the interval $(- x_q, x_q)$, where $x_q \in \left(0,
\frac{\pi}{2}\right)$ is the solution of the transcendental equation $\tan^{-1} x_q =
\frac{1}{q} \left(\frac{\pi}{2} - x_q\right)$. As $q$ increases, $x_q$ decreases from
$x_0 = \frac{\pi}{2}$ to $x_{\infty} = 0$. One finds, for instance, $x_{1/10} \simeq
1.47335$, $x_{1/2} \simeq 1.14446$, and $x_1 \simeq 0.860334$ for the curves
displayed in figure~1.\par
%
%
A similar procedure can be used by taking as inputs the ES potentials known in
quantum mechanics~\cite{levai, cooper} or generalizations thereof. In this way, we can
get in a unified way many of the exact results that have been derived elsewhere by some
ad hoc techniques~\cite{dekar98, dekar99, plastino00, koc02, gonul02a,
alhaidari, dutra00, dutra03,yu, roy, milanovic, plastino98, gonul02b}.\par
%
%
As two new examples, let us mention
\begin{itemize}
\item Kratzer potential
\begin{eqnarray*}
  V_2(x; \gamma) & = & 4 \gamma^2 - \frac{3}{16 z^2(x)} - \frac{\gamma}{z(x)},
         \qquad \gamma > 0, \\
  E_n & = & 4 \gamma^2 - \frac{4\gamma^2}{(2n+1)^2}, \qquad n = 1, 3, 5, \ldots,   
\end{eqnarray*}
\item Symmetric QES potential
\begin{eqnarray*}
  V_2(x; A) & = & \frac{1}{4} A^2 \sinh^2[z(x)] - A \cosh[z(x)] + \frac{1}{2} A +
         \frac{1}{4}, \qquad A > 0, \\
  E_0 & = & 0, \qquad E_1 = A,
\end{eqnarray*}
\end{itemize} 
where we have taken $\nu=0$ for simplicity and we have fixed $\lambda=1$ without any
loss of generality.
\par
%
%
We remark that one of the couplings in Kratzer is fixed at $- \frac{3}{16}$ and so not
free. This entitles it to be classified as a CES potential. Further we observe that the run of
$n$ is restricted to odd values only because physically acceptable wavefunctions then
have the correct $x^{3/4}$ behaviour at the origin (for a discussion of wavefunction
behaviour for strongly singular potentials see~\cite{grosche, znojil} and references
quoted therein). The symmetric QES potential~\cite{tkachuk}, on the other hand, is a
special case of Razavy potential~\cite{razavy} with two known eigenstates. It is needless
to say that broader classes of potentials, such as the ones obtained in~\cite{turbiner,
shifman, ushveridze, bagchi, dutra93, dutt, roychoudhury}, may also be mass deformed
following the approach prescribed in this paper.\par
%
%
\section{Conclusion}

We studied in this Letter a PDEM quantum Hamiltonian in one dimension guided by the
kinetic energy operator $\htt$ of von Roos. The free-particle problem is analyzed first in
the presence of a $\sech^2$-mass background and generation of bound states is noted
by exploiting a first-order intertwining relationship. The non-trivial nature of this result
should be stressed. In a second step, the accompanying potential $V(x)$ of the PDEM
Hamiltonian is found to reduce to two terms --- one involving the ambiguity parameters of
$\htt$ and the other emerging from the coordinate transformation of the constant-mass
Schr\"odinger equation. The advantage of our scheme is that, for a given mass function, a
knowledge of some constant-mass Schr\"odinger potential allows a full access to the
associated potential in the effective-mass Hamiltonian. This provides a unified treatment
of all mass-deformed potentials corresponding to ES, QES or CES potentials known in the
constant-mass case. We illustrated our results by an appropriate choice of the mass
function as applied to the well-known Scarf I potential.\par
%
%
\section*{Acknowledgments}

Two of us, BB and RR, gratefully acknowledge the support of the National Fund for
Scientific Research (FNRS), Belgium, and the warm hospitality at PNTPM, Universit\'e Libre
de Bruxelles, where this work was carried out. PG thanks the Council of Scientific and
Industrial Research (CSIR), New Delhi for the award of a fellowship. CQ is a Research
Director of the National Fund for Scientific Research (FNRS), Belgium.\par
%
%
 
%
%
\newpage
\section*{Figure Caption}

{}Fig.\ 1. Comparison between the potential $V_2(x; 3, 1)$ of Eq.~(\ref{eq:Scarf-q}) for
$q = 0.1$ (dashed line), $q = 0.5$ (dotted line), or $q = 1$ (dot-dashed line) and the
Scarf I potential $U(x; 3, 1)$ (solid line).
%
%
\newpage
\begin{picture}(160,100)
\put(35,0){\mbox{\scalebox{1.0}{\includegraphics{Quesne1.eps}}}}
\end{picture}

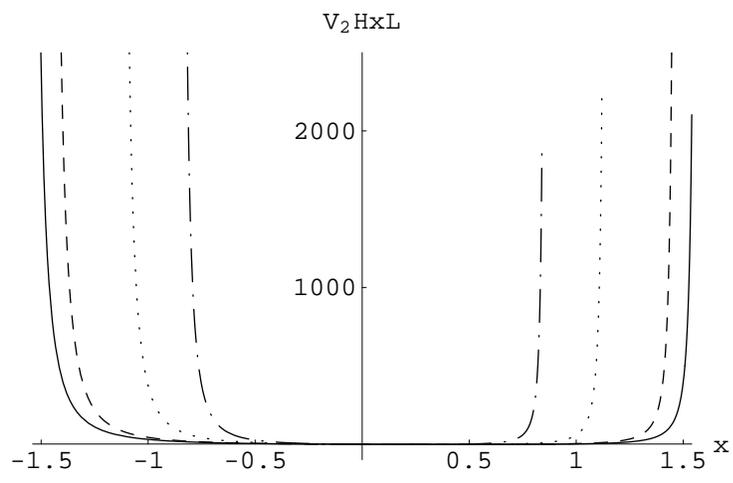
\end{document}